\documentclass[aps,prl,showpacs,twocolumn,floatfix]{revtex4}
\usepackage{amssymb}
\usepackage{graphicx}
\usepackage{dcolumn}
\usepackage{bm}
\usepackage[]{amsmath}

\begin{document}

\title{Coulomb-Enhanced Spin-Orbit Splitting: The Missing Piece in the Sr$_{2}$RhO$_{4}$ Puzzle}
\author{Guo-Qiang Liu, V. N. Antonov, O. Jepsen, and O.K. Andersen.}
\affiliation{Max-Planck Institut f\"ur Festk\"orperforschung, D-70569 Stuttgart, Germany}
\date{\today }

\begin{abstract}
The outstanding discrepancy between the measured and calculated
(local-density approximation) Fermi surfaces in the well-characterized,
paramagnetic Fermi liquid Sr$_{2}$RhO$_{4}$ is resolved by including the
spin-orbit coupling and Coulomb repulsion. This results in an effective
spin-orbit coupling constant enhanced 2.15 times over the bare value. A
simple formalism allows discussion of other systems. For Sr$_{2}$RhO$_{4}$,
the experimental specific-heat and mass enhancements are found to be 2.2.
\end{abstract}

\maketitle

Since the discoveries of high-temperature superconductivity and colossal
magnetoresistance in Mott insulators made metallic by hole-doping,
transition-metal oxides have remained at the forefront of research. Their
many lattice and electronic (orbital, charge, and spin) degrees of freedom
are coupled by effective interactions (electron-phonon, hopping, $t$,
Coulomb repulsion, U, and Hunds-rule coupling, J), and when some of
these are of similar magnitude, competing phases may exist in the region of
controllable compositions, fields, and temperatures. The interactions tend
to remove low-energy degrees of freedom, e.g.\thinspace to reduce the
metallicity. This rarely happens by merely shifting spectral weight from a
quasiparticle band into incoherent Hubbard bands, as in the U/$t$-driven
metal-insulator transition for the single-band Hubbard model, but is usually
assisted by lattice distortions which break the degeneracy of low-energy
orbitals and split the corresponding quasiparticle --or partly incoherent--
bands away from the chemical potential. According to recent calculations
using the local density-functional plus \emph{dynamical} mean-field
approximation (LDA+DMFT), such Coulomb-enhanced crystal-field splitting
seems to be the mechanism triggering the expansion-induced metal-insulator
transition in undoped LaMnO$_{3}$ \cite{LaMnO3} and in V$_{2}$O$_{3}$ \cite%
{V2O3}, long considered the prototype Mott transition. The low-temperature,
antiferromagnetically-ordered, insulating phase of V$_{2}$O$_{3}$ is well
described \cite{Ezhov} in the LDA+U \emph{static} mean-field
approximation, which yields the configuration $t_{2g}^{~2}\rightarrow
e_{g}^{\pi ~\uparrow \uparrow }\,a_{1g}^{~0}.$ Although this approximation
exaggerates the tendency towards symmetry breaking, it does give a
reasonable description of the shape of the Fermi surface (FS) on the
metallic side of the transition \cite{LaMnO3,V2O3}.

When going from 3$d$ to 4$d$ transition-metal oxides, the larger extent of
the 4$d$ orbitals cause the hopping, $t,$ and the coupling to the lattice to
increase, and U and J to decrease. This is reflected in the rich
electronic properties of e.g. the $t_{2g}^{~4}$ ruthenates in the
Ruddlesden-Popper series (Ca$_{1-x}$Sr$_{x}$)$_{\nu +1}$Ru$_{\nu }$O$_{3\nu
+1}$ \cite{Nakatsuji_2,Braden,Nakatsuji_1,Friedt,Friedt06}. Here, the
end-members ($\nu \mathrm{=}1$ and $\nu \mathrm{=}\infty )$ have the same
structures as respectively La$_{2}$CuO$_{4}$ (2D K$_{2}$NiF$_{4}$-type) and
LaMnO$_{3}$ (3D perovskite). The relatively small size and strong covalency
of the Ca ions cause the RuO$_{6}$ octahedra to rotate and tilt. The
resulting misalignment of the Ru $t_{2g}$ Wannier orbitals (WOs) reduces the
hopping between them, and so does the deformation of the WOs caused by Ca-O-$%
t_{2g}$ covalency \cite{Pavarini,Maiti}. As a result, the 2D materials with $%
x\lesssim 0.1$ are Mott-insulators. Ca$_{2}$RuO$_{4}$ is insulating below
360 K, has orbital order with flat octahedra below 260$\,$K \cite{05Keimer},
and is antiferromagnetic below 110$\,$K with configuration $xy^{\,\uparrow
\downarrow }\,xz^{\,\uparrow }\,yz^{\,\uparrow }$ according to the LDA+U 
\cite{Park01,Fang04}. Moderate pressure induces a first order-transition to
a metallic state with reduced tilt and rotation, and with ferromagnetic
order at low temperature \cite{Nakamura02,Snow02,Steffens05}. For $x\gtrsim
0.1$ the materials are metallic and exhibit metamagnetism coupled to
structural distortions as long as $x\leq 0.25$ \cite{05Kriener,07Steffens}.
The properties of the ruthenates with $x\gtrsim 0.1$ seem to be well
described \cite{Oguchi,Singh,Woods,Fang01,Fang04,Maiti05,Singh06} in the
spin-unrestricted LDA (LSDA), a parameter-free approximation which
essentially neglects U, and substitutes J by the Stoner exchange
coupling. For Ca$_{1.5}$Sr$_{0.5}$RuO$_{4}$ at 40$\,$K, angle-resolved
photoemission spectroscopy (ARPES) gives a FS which --after paying due
attention to surface reconstruction-- is found \cite{Wang} to have neither orbital nor
spin-polarization, and to be in good agreement with the LDA.
Finally, stoichiometric Sr$_{2}$RuO$_{4}$ is tetragonal and becomes
superconducting below $1\,$K, presumably with spin-triplet $p$-wave pairing 
\cite{Maeno,RMP}. Both dHvA \cite{Mackenzie} and ARPES \cite%
{Damascelli,Ingle} measurements show that, at low temperature, Sr$_{2}$RuO$%
_{4}$ is a nearly 2D Fermi liquid whose FS agrees well with the LDA \cite%
{Oguchi,Singh} and a mass-enhancement of about 3.

In view of this decreasing strength of the Coulomb correlations, it
therefore came as a surprise when dHvA and ARPES \cite{Perry,Kim,Baumberger}
at $\sim $10$\,$K showed Sr$_{2}$RhO$_{4},$ also a paramagnetic Fermi liquid
with similar structure and electron-electron interactions but one more
electron, to have a FS in substantial disagreement with the LDA. The initial
surprise was that the experimental FS has no $xy$ sheet, but it was soon
realized \cite{Kim,Ko} that in the K$_{2}$NiF$_{4}$ structure the 2D $xy$ and $%
x^{2}-y^{2}$ bands are so broad that they overlap at the $d^{5}$ Fermi
level, and that the observed \cite{Vogt} relaxation by alternating rotations
of the neighboring, corner-sharing RhO$_{6}$ octahedra around their $z$%
-axes, is such as to \emph{gap} those two bands. As a consequence, only the
equivalent 1D $xz$ and $yz$ bands, which hardly hybridize with the other $d$
bands nor with each other, remain at the Fermi level with the single $t_{2g}$
hole distributed equally between them. However, also the LDA FS calculated
for the proper structure deviates substantially from the experimental FS 
\cite{Perry,Kim,Baumberger}. This discrepancy clearly seen in Fig.$\,$\ref%
{FSbars}\thinspace (LDA) is disturbing because there is no experimental
indication of any further distortion. Hence, Coulomb-enhanced crystal-field
splitting can not be the solution to this puzzle.

\begin{figure}[tbp]
{\scalebox{0.7}[0.7]{\includegraphics{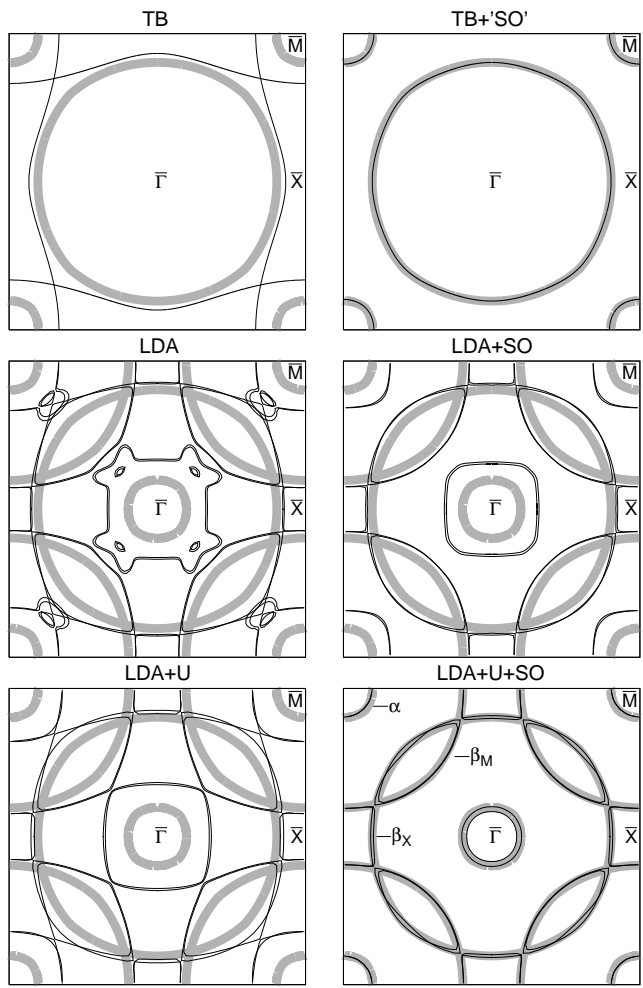}}}
\caption{FS of Sr$_{2}$RhO$_{4}$ in the $k_{z}\mathrm{=}0$ plane of the
large Brillouin zone centered at $\left( k_{x},k_{y}\right) =\left(
0,0\right) \equiv \bar{\Gamma},$ with corners at $\left( \protect\pi ,%
\protect\pi \right) \equiv \mathrm{\bar{M},}$ and edge-midpoints at $\left( 
\protect\pi ,0\right) \equiv \mathrm{\bar{X},}$ in units of the inverse
Rh-Rh nearest-neighbor distance. \textit{Grey:} ARPES \protect\cite%
{Kim,Baumberger}, the same FS in all six pictures, but unfolded in the top
panels. \textit{Black:} Theory using six different approximations. The top
panels results from the analytical expressions and parameter values given in
the text and employs one $xz$ and one $yz$ orbital per cell. The remaining
pictures result from all-orbital LDA\protect\cite{LDA} calculations for the
proper crystal structure with 4 formula units per cell \protect\cite{Vogt}.
The conventional Fermi-sheet notation is given in the (LDA+U+SO) picture.}
\label{FSbars}
\end{figure}

\begin{figure}[tbp]
{\scalebox{0.48}[0.48]{\includegraphics{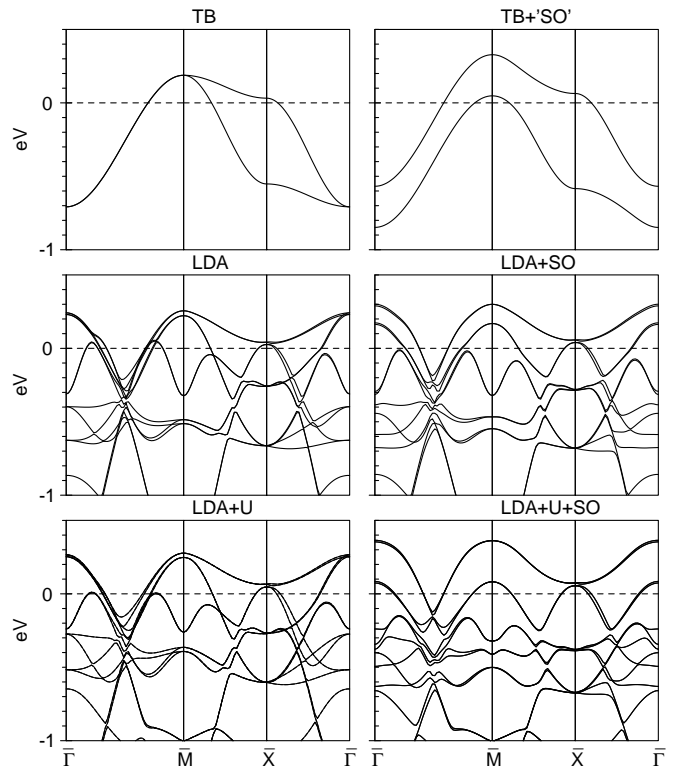}}}
\caption{Theoretical band structure of Sr$_{2}$RhO$_{4}$ in the six
different approximations. See Fig. 1.}
\label{BND_n}
\end{figure}

In this Letter we shall argue that the clue is \emph{Coulomb-enhanced
spin-orbit (SO) splitting.} We begin by demonstrating that the experimental FS can be
perfectly fitted and the low-energy band structure explained using simple,
analytical theory. These results are illustrated in the top panels of Fig.s 
\ref{FSbars} and \ref{BND_n}. We then perform $ab$ $initio$ LDA
calculations \cite{LDA} including the SO coupling (LDA+SO), and later also
on-site Coulomb effects in the LDA+U approximation. The detailed results
in the 2nd and 3rd panels of Fig.s \ref{FSbars} and \ref{BND_n} prove the
soundness of our simple theory.

In the Hilbert space of only the $xz$ and $yz$ WOs, the 2D translational
symmetry is a quadratic lattice with \emph{one} Rh per cell; the relative
rotation of the nearest-neighbor RhO$_{6}$ octahedra merely modifies the
hopping integrals \cite{Pavarini}. In the $z$-direction the hopping is small
and not detected in ARPES. Neglecting it, as well as the small
rotation-induced hopping between the $xz$ and $yz$ bands, the electron
dispersion is \cite{Pavarini,PavMazin}:%
\begin{equation}
\varepsilon _{xz}^{\mathbf{k}}=-2t_{\pi }\cos k_{x}-2t_{\delta }\cos
k_{y}-\varepsilon _{F}\,,  \label{TB}
\end{equation}%
and equivalently for $\varepsilon _{yz}^{\mathbf{k}}.$ Here, $\varepsilon
_{F}$ is the position of the Fermi level with respect to the energy at $\bar{%
\Gamma}\mathrm{\bar{M}}/2$ and signs have been chosen such that the hopping
integrals are positive: $t_{\pi }\sim t_{pd\pi }^{2}/\left( \varepsilon
_{F}-\epsilon _{p}\right) $ because this hop is mainly via the O$\,p_{z}$
orbital, and $0<t_{\delta }\ll t_{\pi }$ because this hop is direct $%
dd\delta .$ These tight-binding bands are shown in Fig.$\,$\ref{BND_n}%
\thinspace (TB). In the space of the $\left( xz\uparrow ,\,xz\downarrow
,\,yz\uparrow ,\,yz\downarrow \right) $ WOs, the eigenfunctions of the SO
coupling are:%
\begin{equation*}
\chi _{m_{j}=\frac{3}{2}}=\left( xz+iyz\right) \uparrow \;%
\mathrm{and}\;\chi _{-\frac{3}{2}}=\left( xz-iyz\right)
\downarrow 
\end{equation*}%
with eigenvalue $\frac{1}{2}\zeta $ $\left( >0\right) ,$ and 
\begin{equation*}
\chi _{\frac{1}{2}}=\left( xz+iyz\right) \downarrow \;\mathrm{and%
}\;\chi _{-\frac{1}{2}}=\left( xz-iyz\right) \uparrow 
\end{equation*}%
with eigenvalue $-\frac{1}{2}\zeta .$ SO coupling thus splits the degeneracy
of the $xz$ and $yz$ bands along ${\bar{\Gamma}} \mathrm{\bar{M}}$ by $\zeta $ as
seen in Fig.$\,$\ref{BND_n}\thinspace (TB+'SO'). Since the structure has
inversion symmetry, all bands remain doubly degenerate. The band structure
is given by:%
\begin{equation}
\varepsilon _{\pm }^{\mathbf{k}}=\frac{1}{2}\left[ \varepsilon _{xz}^{%
\mathbf{k}}+\varepsilon _{yz}^{\mathbf{k}}\pm \sqrt{\left( \varepsilon
_{xz}^{\mathbf{k}}-\varepsilon _{yz}^{\mathbf{k}}\right) ^{2}+\zeta ^{2}}%
\right] ,  \label{TBSO}
\end{equation}%
and the FS, $\varepsilon _{\pm }^{\mathbf{k}}\mathrm{=}0,$ is determined by
merely 3 parameters, $t_{\pi }/\varepsilon _{F},$ $t_{\delta }/\varepsilon
_{F},$ and $\zeta /\varepsilon _{F},$ which may be obtained by fitting to 3
FS dimensions, e.g.\thinspace the two intersections along $\bar{\Gamma}%
\mathrm{\bar{M}}$ and the one along $\bar{\Gamma}\mathrm{\bar{X}.}$ Fitting
in this way to the ARPES FS \cite{Baumberger}, \emph{perfect agreement} with
the \emph{entire} FS is obtained as seen in Fig.$\,$\ref{FSbars}\thinspace
(TB+'SO'). The area of the $\bar{\Gamma}$-centered electron sheet \emph{minus%
} that of the $\mathrm{\bar{M}}$-centered hole sheet equals half the
Brillouin-zone area. As seen in Fig.$\,$\ref{FSbars}\thinspace (LDA+SO), the
agreement with ARPES is less good for the ab initio relativistic LDA FS: the
SO splitting along $\bar{\Gamma}\mathrm{\bar{M}}$ is too small. In fact,
fitting the LDA+SO FS to our analytical expression, yields essentially the
same values for $t_{\pi }/\varepsilon _{F}$ and $t_{\delta }/\varepsilon
_{F},$ but\ $\zeta /\varepsilon _{F}$\ is smaller than $\zeta _{\mathrm{ARPES%
}}/\varepsilon _{F}$ by the \emph{large factor} $2.15.$ The LDA+SO
calculation (Fig.\thinspace \ref{BND_n}) yields: $\zeta =0.13\,$eV, a value
which is smaller than the $0.16\,$eV obtained for elemental fcc Rh \cite%
{MackAn} due to the O$\,p_{z}$ tails of the rhodate WOs. Letting $\zeta $
set the energy scale of bare bands, the TB+'SO' parameters reproducing the
ARPES FS are: $\zeta _{\mathrm{ARPES}}=2.15\times 0.13\,\mathrm{eV}=0.28\,$%
eV, $\varepsilon _{F}=0.260$ eV, $t_{\pi }=0.185$ eV and $t_{\delta }=0.039$
eV. The TB results in Fig.s \ref{FSbars} and \ref{BND_n} are obtained with $%
\zeta =0$.

In order to explain the origin of the SO\emph{-enhancement,} we need to add
the on-site Coulomb repulsion, $\hat{H}_{\mathrm{C}}^{\mathbf{n}},$ and thus
consider the 2-band Hubbard Hamiltonian:%
\begin{equation}
\hat{H}=\sum_{\mathbf{k}}\sum_{\mu =xz,yz}\sum_{\sigma }\varepsilon _{\mu }^{%
\mathbf{k}}\hat{c}_{\mu \sigma }^{\mathbf{k~}\dagger }\hat{c}_{\mu \sigma }^{%
\mathbf{k}}+\sum_{\mathbf{n}}\left( \hat{H}_{\mathrm{SO}}^{\mathbf{n}}+\hat{H%
}_{\mathrm{C}}^{\mathbf{n}}\right) ,  \label{H}
\end{equation}%
$\varepsilon _{\mu }^{\mathbf{k}}$ given by Eq.$\,$(\ref{TB}) and $\mathbf{%
n\equiv }\left( n_{x},n_{y}\right) $ runs over the quadratic lattice. For
simplicity, we drop the superscripts $\mathbf{n}$ in the following. The SO
interaction is:%
\begin{equation}
\hat{H}_{\mathrm{SO}}=\frac{\zeta }{2}\left( \hat{n}_{\frac{3}{2}%
}+\hat{n}_{-\frac{3}{2}}-\hat{n}_{\frac{1}{2}}-\hat{n%
}_{-\frac{1}{2}}\right) \equiv -\frac{\zeta }{2}\hat{p},
\label{SO}
\end{equation}%
where $\hat{n}_{m_{j}}$ is an electron-number operator and $\hat{p}$ is
the difference between $|m_{j}|\mathrm{=}\frac{3}{2}$ and $%
|m_{j}|\mathrm{=}\frac{1}{2},$ i.e.: the SO polarization. The Coulomb repulsion is 
U between two electrons with the same orbital, U-2J between electrons
with different orbitals and different spins, and U-3J between electrons
with different orbitals:%
\begin{eqnarray}
\hat{H}_{\mathrm{C}} &=&U\left( \hat{n}_{\frac{3}{2}}\hat{n}_{%
\frac{1}{2}}+\hat{n}_{-\frac{3}{2}}\hat{n}_{%
-\frac{1}{2}}\right) \notag \\   
&&+\left( U-2J\right) \left( \hat{n}_{\frac{3}{2}}\hat{n}_{%
-\frac{3}{2}}+\hat{n}_{\frac{1}{2}}\hat{n}_{-%
\frac{1}{2}}\right)   \notag \\
&&+\left( U-3J\right) \left( \hat{n}_{\frac{3}{2}}\hat{n}_{%
-\frac{1}{2}}+\hat{n}_{-\frac{3}{2}}\hat{n}_{%
\frac{1}{2}}\right).   \notag
\end{eqnarray}%
Spin-spin correlations have been neglected here and there is no
double-counting correction because we consider equivalent orbitals. Since Sr$%
_{2}$RhO$_{4}$ is paramagnetic at low temperature, we can set $\hat{n}_{%
\frac{3}{2}}=\hat{n}_{-\frac{3}{2}}\equiv \frac{1}{4}%
\left( \hat{n}-\hat{p}\right) $ and $\hat{n}_{\frac{1}{2}}=\hat{n%
}_{-\frac{1}{2}}\equiv \frac{1}{4}\left( \hat{n}+\hat{p}\right) $
with $\hat{n}$ being the number of electrons in the two doubly-degenerate
bands, and get:%
\begin{equation*}
\hat{H}_{\mathrm{C}}\approx \frac{3U-5J}{8}\hat{n}^{2}-\frac{U-J}{8}\hat{p}%
^{2}\approx c-\frac{U-J}{4}p\hat{p}.
\end{equation*}%
The last expression is the mean-field approximation, which grouped together
with $\hat{H}_{\mathrm{SO}}$ (\ref{SO}) yields a one-electron Hamiltonian
with SO-coupled bands (\ref{TBSO}), but with $\zeta $ substituted by $\zeta
_{eff}=\zeta +\frac{1}{2}\left( U-J\right) p,$ and $p$ determined
self-consistently. The polarization function, $p\left( \zeta \right) ,$ may
by found from the polarization of each Bloch state,%
\begin{equation*}
\left\vert c_{\pm ;\frac{3}{2}}^{\mathbf{k}}\right\vert ^{2}-\left\vert
c_{\pm ;\frac{1}{2}}^{\mathbf{k}}\right\vert ^{2}=\frac{\partial \varepsilon
_{\pm }^{\mathbf{k}}}{\partial \zeta /2}=\frac{\pm \zeta }{\sqrt{\left(
\varepsilon _{xz}^{\mathbf{k}}-\varepsilon _{yz}^{\mathbf{k}}\right)
^{2}+\zeta ^{2}}},
\end{equation*}
as follows from 1st-order perturbation theory. As a result:%
\begin{equation*}
p\left( \zeta \right) =2\zeta \int_{0}^{\pi }\int_{0}^{\pi }\frac{%
dk_{x}dk_{y}}{\pi ^{2}}\frac{\theta \left( \varepsilon _{+}^{\mathbf{k}%
}\right) \theta \left( -\varepsilon _{-}^{\mathbf{k}}\right) }{\sqrt{\left(
\varepsilon _{xz}^{\mathbf{k}}-\varepsilon _{yz}^{\mathbf{k}}\right)
^{2}+\zeta ^{2}}}.
\end{equation*}%
where the factor 2 takes the double degeneracy of each band into account and
we have used that $\theta \left( -\varepsilon _{-}^{\mathbf{k}}\right)
-\theta \left( -\varepsilon _{+}^{\mathbf{k}}\right) =\theta \left(
\varepsilon _{+}^{\mathbf{k}}\right) \theta \left( -\varepsilon _{-}^{%
\mathbf{k}}\right) .$ The integral is over the area between the $\bar{\Gamma}
$-centered electron sheet and the $\mathrm{\bar{M}}$-centered hole sheet$.$
For $\zeta $ large, $p\left( \zeta \right) \rightarrow 4-n$, where $n$ is the number of 
electrons in the $xz$ and $yz$ bands,
and for $\zeta $ small, $p\left( \zeta \right) \rightarrow
\chi _{0}\zeta $ with the susceptibility%
\begin{equation}
\chi _{0}=2\int_{0}^{\pi }\int_{0}^{\pi }\frac{dk_{x}dk_{y}}{\pi ^{2}}\frac{%
\theta \left( \varepsilon _{+}^{\mathbf{k}}\right) \theta \left(
-\varepsilon _{-}^{\mathbf{k}}\right) }{\varepsilon _{xz}^{\mathbf{k}%
}-\varepsilon _{yz}^{\mathbf{k}}}\approx \frac{1}{2\left( t_{\pi }-t_{\delta
}\right) },  \label{chi}
\end{equation}%
which turns out to be independent of $n$.
The last approximation results from integrating over the rectangular $%
t_{\delta }\mathrm{=}0$ area. The self-consistency condition, $\zeta
_{eff}=\zeta +\frac{1}{2}\left( U-J\right) p\left( \zeta _{eff}\right) ,$
thus yields a \emph{Coulomb-enhanced SO coupling, }which for $\zeta _{eff}$
in the \emph{linear} range of the polarization function is given by:%
\begin{equation}
\frac{\zeta _{eff}}{\zeta }\approx \left[ 1-\frac{U-J}{2}\chi _{0}\right]
^{-1}\approx \left[ 1-\frac{U-J}{4\left( t_{\pi }-t_{\delta }\right) }\right]
^{-1}.  \label{SOenh}
\end{equation}%
Inserting $\zeta _{eff}/\zeta =2.15$ and $t_{\pi }-t_{\delta }=0.146$%
\thinspace eV, we get: $ U-J=0.3eV$ and $p\left( \zeta _{eff}\right)
=\allowbreak 0.97,$ whereas using the proper polarization function, which
saturates at $p\left( \infty \right) \mathrm{=}1$, yields: $ U-J=0.5eV$.
This is a reasonable value for a 4$d$ WO spreading onto the oxygen sites. 

In order to substantiate this simple picture, we perform all-orbital
relativistic LDA+U calculations \cite{LDA} with U-J adjusted such as to
give the best agreement with the ARPES FS. Since U and J in such calculations do not refer to proper orbitals, but to
d-waves truncated and normalized inside atomic (LMTO) or muffin-tin (LAPW) spheres, the values of the parameters
depend on the sphere size and are generally larger than for the more diffuse WOs \cite{LDA}. As is obvious from Fig.\thinspace %
\ref{FSbars}\thinspace (LDA+U+SO), this agreement is even more perfect than
for TB+'SO'; now even the small observed gaps \cite{xz/yzHyb} induced by the
AF rotations are reproduced. The LDA+U+SO bands in Fig.\thinspace \ref{BND_n}
are well reproduced in the range from 0.15 eV below- to 0.5 eV above the
Fermi level by the TB+'SO' bands folded into the BZ/2 with corners at $%
\mathrm{\bar{X}.}$ Features not reproduced are the tiny splittings due to
in-plane $xz$-$xy$ hopping and out-of-plane hoppings neglected in our TB
model. The agreement between the TB and the LDA calculation is less
satisfactory, first of all because without SO-quenching the in-plane $xz$-$xy
$ hopping produces a splitting along $\bar{\Gamma}\mathrm{\bar{M},}$ and
secondly because the rotation-induced $xy$-$\left( x^{2}-y^{2}\right) $
gapping is not complete without SO coupling. In fact, it takes the LDA+U+SO
to push the lower edges of the $xy$-$\left( x^{2}-y^{2}\right) $ gap to $%
-0.16$\thinspace eV along $\bar{\Gamma}\mathrm{\bar{M}}$, and even deeper
along $\bar{\Gamma}\mathrm{\bar{X}}$, locations close to those observed with
ARPES \cite{Perry}. Note finally, that the LDA+U alone, without SO coupling,
brings little improvement compared with the LDA.

Having obtained perfect agreement with the ARPES FS, we use the LDA+U+SO to
calculate cyclotron masses and compare with those obtained from the dHvA
measurements \cite{Perry}. The resulting enhancements are: $m_{\mathrm{ARPES}%
}/m_{\mathrm{LDA+U+SO}}=2.1\left( \alpha \right) ,~2.1\left( \beta _{\mathrm{%
M}}\right) ,$ and $2.3\left( \beta _{\mathrm{X}}\right) .$ Consistently
herewith the density of states at the Fermi level yields the electronic
specific-heat enhancement: $\gamma _{\exp }/\gamma _{\mathrm{LDA+U+SO}}=2.2.$
These many-body enhancements are smaller than those $\left( \sim 3\right) $
in Sr$_{2}$RuO$_{4}.$

We have finally performed relativistic LDA+U calculations also for Sr$_{2}$%
RuO$_{4}$, and find the agreement with the experimental FS \cite%
{Mackenzie,Damascelli,Ingle} to improve from good to perfect. Inclusion of
SO+U in the LDA reduces the area of the $\alpha $-pocket from 13 to 11\% of
the BZ-area in Sr$_{2}$RuO$_{4}$, as compared with a 24 to 6\% reduction in
Sr$_{2}$RhO$_{4}$. The stronger SO effects in the latter oxide are caused by 
$\zeta $ being 20\% larger \cite{MackAn}, due to the larger mass, and by the
Coulomb enhancement (\ref{SOenh}) being larger due to the reduction of $%
t_{\pi }$ caused by rotation. Other factors are similar in the two oxides
because the extra hole in Sr$_{2}$RuO$_{4}$ is accomodated in the $xy$ band,
which hardly couples with the $xz$ and $yz$ bands. It is conceivable that
reduction of $t_{\pi }$ caused by Ca-induced rotations could make
Coulomb-enhanced SO coupling of the $xz$ and $yz$ bands important in other
ruthenates and rhodates not driven magnetic by the $t$-reduction.

In conclusion, resolution of the Sr$_{2}$RhO$_{4}$ puzzle has tought us that
although usually neglected in $4d$-oxides, the spin-orbit coupling belongs
to the list of competing interactions which cause the rich physics of these
materials \cite{Zawa}.


\begin{thebibliography}{99}
\bibitem{LaMnO3} A. Yamasaki \textit{et al., }Phys. Rev. Lett. \textbf{96},
166401 (2006).

\bibitem{V2O3} A.I. Poteryaev \textit{et al.}, Phys. Rev. B \textbf{76},
085127 (2007).

\bibitem{Ezhov} S. Y. Ezhov \textit{et al.}, Phys. Rev. Lett. \textbf{83},
4136 (1999).

\bibitem{Nakatsuji_2} S. Nakatsuji \textit{et al.}, Phys. Soc. Jpn \textbf{66%
}, 1868 (1997).

\bibitem{Braden} M. Braden $et$ $al.$, Phys. Rev. B \textbf{58} , 847 (1998).

\bibitem{Nakatsuji_1} S. Nakatsuji \textit{et al.,} Phys. Rev. Lett. \textbf{%
84}, 2666 (2000).

\bibitem{Friedt} O. Friedt $et$ $al.$, Phys. Rev. B \textbf{63}, 174432
(2001);

\bibitem{Friedt06} O. Friedt $et$ $al.$, Phys. Rev. Lett. \textbf{93} ,
147404 (2004).

\bibitem{Pavarini} E. Pavarini \textit{et al.}, New J. Phys. \textbf{7}, 188
(2005).

\bibitem{Maiti} K. Maiti, Phys. Rev. B \textbf{73}, 235110 (2006)

\bibitem{05Keimer} I. Zegkinoglou \textit{et al.}, Phys. Rev. Lett. \textbf{%
95}, 136401 (2005).

\bibitem{Park01} K. T. Park, J. Phys. Condens. Matter \textbf{13}, 9231
(2001)

\bibitem{Fang04} Z. Fang \textit{et al.}, Phys. Rev. B \textbf{69}, 045116
(2004)

\bibitem{Nakamura02} F. Nakamura \textit{et al.}, Phys. Rev. B \textbf{65},
220402(R) (2002).

\bibitem{Snow02} C.S. Snow \textit{et al.}, Phys. Rev. Lett. \textbf{89},
226401 (2002).

\bibitem{Steffens05} P. Steffens \textit{et al.}, Phys. Rev. B \textbf{72},
094104 (2005).

\bibitem{05Kriener} M. Kriener \textit{et al.}, Phys. Rev. Lett. \textbf{95}%
, 267403 (2005).

\bibitem{07Steffens} P. Steffens \textit{et al.}, Phys. Rev. Lett. \textbf{99%
}, 217402 (2007).

\bibitem{Oguchi} T. Oguchi, Phys. Rev. B \textbf{51}, 1385(R) (1995).

\bibitem{Singh} D. J. Singh, Phys. Rev. B \textbf{52}, 1358 (1995).

\bibitem{Woods} L. M. Woods, Phys. Rev. B \textbf{62}, 7833 (2000)

\bibitem{Fang01} Z. Fang $et$ $al.$, Phys. Rev. B \textbf{64}, 020509(R)
(2001)

\bibitem{Maiti05} K. Maiti $et$ $al.$, Phys. Rev. B \textbf{71}, 161102(R)
(2005)

\bibitem{Singh06} D.J. Singh $et$ $al.$, Phys. Rev. Lett. \textbf{96},
097203 (2006)

\bibitem{Wang} S.-C. Wang $et$ $al.$, Phys. Rev. Lett. \textbf{93}, 177007
(2004).

\bibitem{Maeno} Y. Maeno $et$ $al.$, Nature (London) \textbf{372}, 532
(1994).

\bibitem{RMP} A.P. Mackenzie $et$ $al.$, Rev. Mod. Phys. \textbf{75}, 657
(2003)

\bibitem{Mackenzie} A.P. Mackenzie $et$ $al.$, Phys. Rev. Lett. \textbf{76},
3786 (1996).

\bibitem{Damascelli} A. Damascelli $et$ $al.$, Phys. Rev. Lett. \textbf{85},
5194 (2000).

\bibitem{Ingle} N.J.C. Ingle \textit{et al.}, Phys. Rev. B \textbf{72}, 205114
(2005)

\bibitem{Kim} B. J. Kim $et$ $al.$, Phys. Rev. Lett. \textbf{97}, 106401
(2006).

\bibitem{Perry} R.S. Perry $et$ $al.$, New J. Phys. \textbf{8}, 175 (2006).

\bibitem{Baumberger} F. Baumberger $et$ $al.$, Phys. Rev. Lett. \textbf{96},
246402 (2006).

\bibitem{Ko} E. Ko $et$ $al.$, Phys. Rev. Lett. \textbf{98}, 226401 (2007).

\bibitem{Vogt} T. Vogt and D. J. Buttrey, J. Solid State Chem. \textbf{123},
186 (1996); $T\mathrm{=}4\,$K, space group: $I4_{1}/acd.$ Compared with the
bct K$_{2}$NiF$_{4}$-type structure, the cell is doubled in the plane due to
the alternating rotations of neighboring RhO$_{6}$ octaheda, and also in the 
$z$-direction due to alternating rotations in every 3rd RhO$_{2}$ plane.

\bibitem{LDA} The calculations were performed for the correct structure \cite%
{Vogt} using RLAPW software for accuracy (K. Schwarz $et$ $al.$, Comput.
Phys. Commun. \textbf{147}, 71 (2002)) and RLMTO software for speed and
intrepretation (V. N. Antonov $et$ $al.$, J. Magn. Magn. Mater. {\bf 146}, 205 (1995)). 
The space filling spheres of the latter method
were chosen such that the bands were identical with those from the former.
The LDA results shown in the figures agree with those previously published 
\cite{Kim,Baumberger}. The LDA+U(+SO) (R)LMTO calculations used U=
2.5eV and J=0.9eV for the Rh $d$ partial waves in the Rh
sphere.

\bibitem{xz/yzHyb} FIG. 3 in \cite{Baumberger}, TABLE I in \cite{PavMazin}
and p.$\,$17 in \cite{Pavarini}.

\bibitem{PavMazin} E. Pavarini and I.I. Mazin, Phys. Rev. B \textbf{74},
035115 (2006); \textit{ibid }\textbf{76}, 079901(E) (2007). The 500$\,$K
scale in Fig.$\,$2 is several times too big.

\bibitem{MackAn} A.R. Mackintosh and O.K. Andersen in \textit{Electrons at
the Fermi Surface, }ed. M. Springford, CUP, Cambridge 1980.

\bibitem{Zawa} Important SO effects in Ca-ruthenates were anticipated by T.
Mizokawa \textit{et al.}, Phys. Rev. Lett. \textbf{87}, 077202 (2001) and Z.
Fang \textit{et al.}, New J. Phys. \textbf{7}, 66 (2005).
\end{thebibliography}
\end{document}